# The super resonance effect paves the way for a new type of refractive index sensor concept based on a mesoscale dielectric sphere

Oleg V. Minin[a], Igor V. Minin[a], and Song Zhou[b]
[a]Tomsk Polytechnic University, 30 Lenin Ave., Tomsk 634050 Russia
[b]Jiangsu Key Laboratory of Advanced Manufacturing Technology, Faculty of Mechanical and Material Engineering, Huaiyin Institute of Technology, Huai'an 223003, China.

**Abstract**

Recently, we showed that dielectric mesoscale spheres could support so-called super-resonance effect, i.e. high-order Mie resonance modes with giant field localization and enhancement. Due to the presence of the surrounding medium leads to a significant influence in the intensity of the field in the particle, based on Mie theory we show for the first time that this effect may be use to design refractive index sensor of medium. Using the example of air as an environment, we have shown that the sensitivity of the proposed sensor concept can reach from $10^{-6}$ to $10^{-8}$, depending on the accuracy of the sphere size parameter, which is no worse than the accuracy of modern interference methods.

**Introduction**

The most common method for measuring the refractive index of a gaseous medium, such as air, are various interferometric methods and their modifications [1-8]. In these methods, the shift of the interference pattern is proportional to the change in the refractive index of the substance. The accuracy of measuring the refractive index of air is determined primarily by the accuracy of the length of the vacuum resonator and the ellipticity of the polarization components of a laser and the accuracy of air refractive index measurement is about from $10^{-6}$ [3] to $5.0 \times 10^{-8}$ [1,6].

Another class of methods consists in measuring the parameters of air composition (air pressure, ambient temperature, humidity, and so on) and calculating the refractive index of air using semi-empirical formulas [9–11]. At the same time, the key and important characteristic of such indirect measurement methods is the simplicity of its application.

Mesoscale dielectric spherical particles did not arouse close interest for a long time, however, it was precisely when the particle radius is on the order of several wavelengths that a whole series of new optical phenomena were discovered [12,13].

For example, in 2019, we reported that lossless dielectric mesoscale spheres, immersed in vacuum (refractive index of medium equal to 1) can support so-called 'super-resonance modes' [14-16], which is different from other types of resonances [17-20] in such particles. This super resonance is associated with high-order internal Mie modes and take place for specific values of the particle size parameter $q$ (defined as $q=2\pi R/\lambda$, where $R$ is the radius of particle and $\lambda$ the incident wavelength), which can be directly obtained from rigorous analytical Mie theory [21].

Recall that the formulas of the Mie theory represent expansions of electromagnetic fields in terms of spherical harmonics, the expansion coefficients of which satisfy the boundary conditions for the continuity of the tangential components of the fields on the particle surface. In the case of a dielectric sphere, the high-order internal resonance mode interferes with a wide spectrum of all other modes. In [14] it was shown that, under certain conditions, a single term with a high-order resonant mode can lead to a multiple increase in the scattered electric and magnetic fields intensity. At the same time, such sharp resonant modes are very sensitive to the Mie size parameter $q$ [14,18].

Recently, based on the Mie theory, it was shown for the first time that the presence of a medium around a particle, for example, air instead of vacuum, leads to a significant decrease in the field intensity enhancement in the particle and a shift in the position of the resonance [22]. Thus, super-resonances modes are also extremely sensitive to the refractive index of the environment.
Below we propose a new physical concept for indirect measurement of the refractive index of a substance (using the example of air), based on the super resonance effect.

**Super-resonance effect vs refractive index of surrounding medium**

In our previous studies [14-16] we have been using a size parameter $q$ sampling accuracy of $\delta q = 10^{-4}$. However, the discretization accuracy of the size parameter affects the accuracy of finding super-resonant modes and the maximum achievable fields intensity enhancement characteristics [18]. Thus, an increase in the discretization accuracy of the size parameter from $10^{-4}$ to $10^{-10}$ leads to an increase in intensity near the particle boundary by 5 orders of magnitude [18]. Accordingly, the width of the resonance peak also changes. Therefore, below we will consider two examples of the potential accuracy of determining the refractive index of a medium around a sphere with a size localization accuracy of $10^{-4}$ and $10^{-10}$.

As a particle material, we select glass with refractive index equal to $n_s = 1.9$, which is near but less than 2 [23,24]. The accurate refractive index of air we select as n=1.000241307.

Once super-resonance peaks position and number of modes for selected values of $q$ and refractive index of the medium were found, we simulate the magnetic and electric fields intensity distribution in XZ plane (incident beam x-polarized, propagates from -z to +z direction). The simulation was performed with spatial resolution $a/200$ within XZ plane ranging from *-1.2a to 1.2a*, where *a* is radius of particle.

For super-resonant mode two characteristic almost symmetrical hotspots both for magnetic and electric fields appear in the illuminated and shadow particle hemispheres along the light propagation direction (the strong enhancement of the electromagnetic field near the back and forward directions close to the surface of spherical particles [18]), showing such modes are strongly coupled to a circulation mode within the sphere [14].

In the Figure 1, we illustrate the super-resonance effect for the non-absorbing mesoscale particle with size parameter q=21.8401542641 and refractive index $n_s$=1.9, immersed in air with refractive index of n=1.000241307. These parameters correspond to a resonant mode excited inside the particle with partial wave order *l*=35.

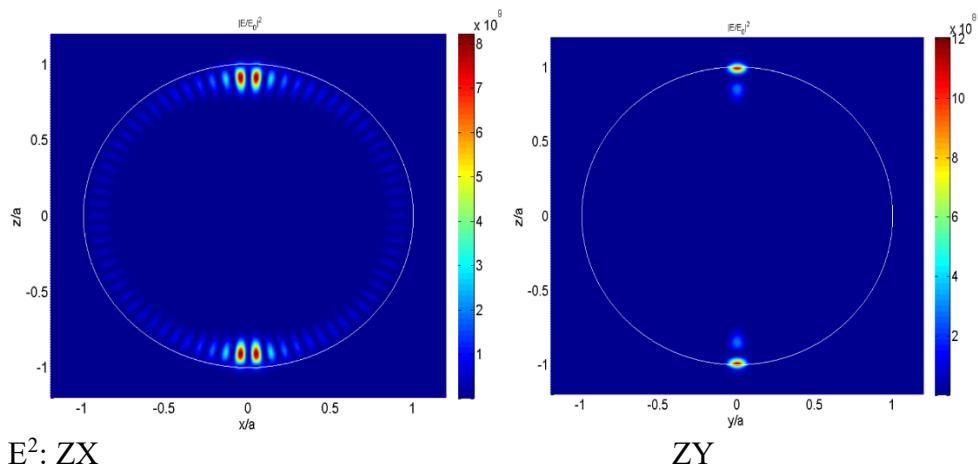

E²: ZX  ZY

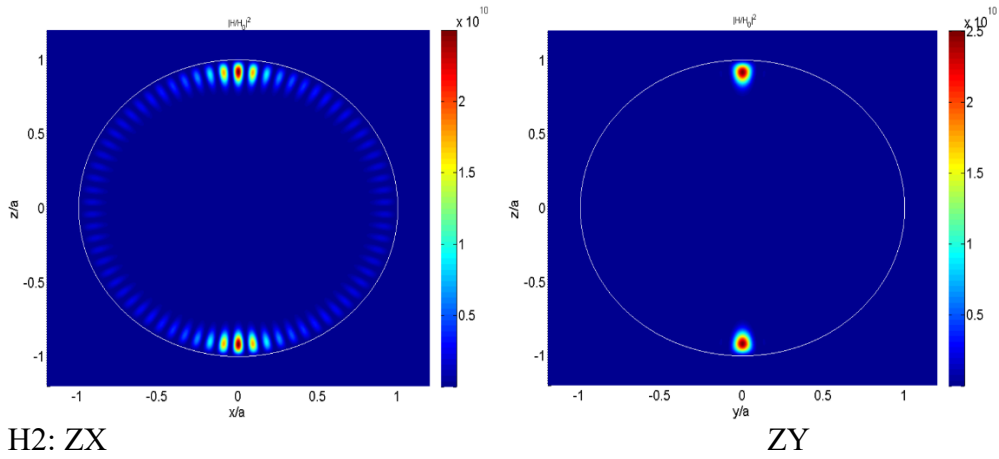

H2: ZX                                                                          ZY

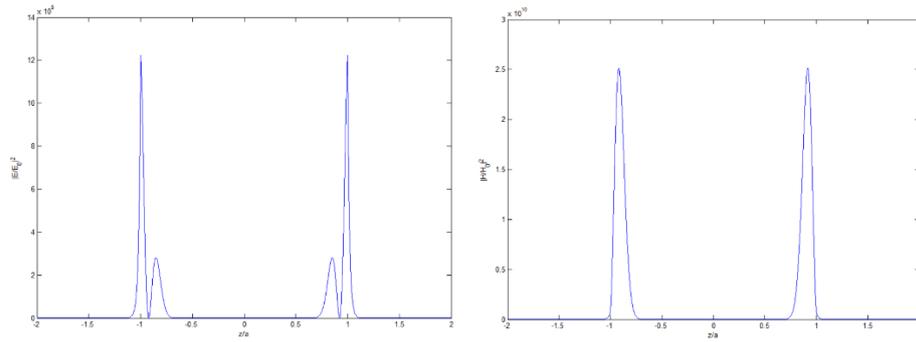

Figure 1. Electric and magnetic hot spots generation for dielectric sphere immersed in air. The distribution of the electric field intensity along the hot spots (along the vertical axis) is shown below.

One can see the maximum electric field intensity enhancement reaches $|E|^2=1.225e9$ and magnetic field about 20 times more - $|H|^2=2.511e10$.

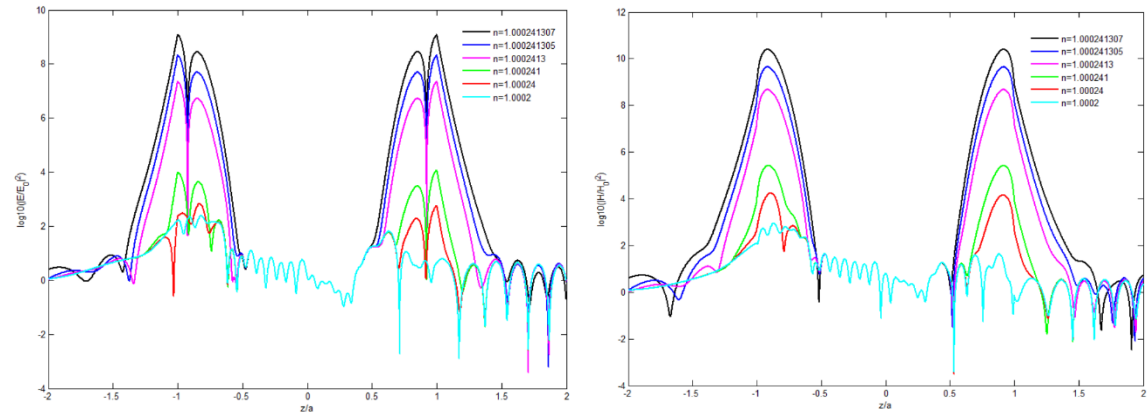

Figure 2. Evolution for electric (left) and magnetic (right) field intensity vs the accuracy of the value of the refractive index of air in the environment.

In the Figure 2, we illustrate the super-resonance effect evolution for such mesoscale particle vs variation of the refractive index of surrounding medium. We observed new super-resonance modes effect that are not seen in previous research, with record sensitivity of the resonant peak to very small changes in the refractive index of the medium. One can see the small changes in refractive index of air $\delta n=2*10^{-9}$ (from $n=1.000241307$ to $n=1.000241305$) leads to a significant, almost in order, change in the field intensity at hot spots – from $|E|^2=1.225e9$ to $|E|^2=2.157e8$ and from $|H|^2=2.511e10$ to $|H|^2=4.427e9$, respectively. Other values of peak intensities are given in Table 1 and correspondent fields intensities distributions are given in the Supplement materials.

Table 1.

| n | L, mode | $E^2$(along z-axis) | $H^2$(along z-axis) |
|---|---|---|---|
| 1.000241307 | 35 | 1.225e9 | 2.511e10 |
| 1.000241305 | 35 | 2.157e8 | 4.427e9 |
| 1.0002413 | 35 | 2.303e7 | 4.744e8 |
| 1.000241 | 35 | 1.231e4 | 2.669e5 |
| 1.00024 | 35 | 683.3 | 1.78e4 |
| 1.0002 | 35 | 242.5 | 906.4 |

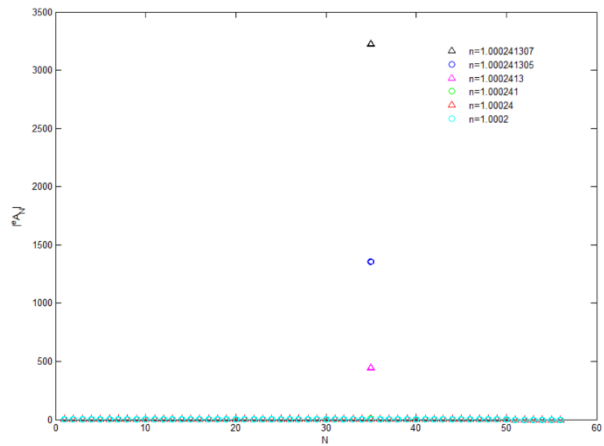

Figure 3. The first 56 Mie $\left|{}^m A_n\right|$ coefficients [22] vs different values of air $n$. Evolution of the partial wave order $l=35$.

The catastrophic drop in intensity is explained by the extremely high sensitivity of the resonant Mie coefficients to the value of the refractive index of the medium [22]. It is well known from the Mie theory that the behavior of the electric and magnetic fields inside the sphere is fully determined by the partial wave scattering amplitudes $c_l$ (TM mode) and $d_l$ (TE mode) [25-27]. The resonances occur at zeros of denominators of $c_l$ and $d_l$. In Figure 3, we show the magnitude of the Mie scattering coefficients $\left|{}^m A_n\right|$ [22] for $1 \leq l \leq 56$ and for different values of $n$. One can see that the resonant scattering coefficient $\left|{}^m A_{35}\right|$ is much higher in magnitude and very sensitive to changes in the refractive index of the medium $n$. A decrease in intensity with a change in the environmental index leads to a change in the conditions of constructive interference of the one partial wave inside the wavelength-scaled sphere, which accordingly affects the abnormal change in magnitude one of the resonant scattering coefficient in the Mie theory than the other coefficients.

Now, let us briefly consider the influence of the refractive index of air environment on the super-resonance effect for a sphere with reduced Mie size parameter sampling accuracy of $\delta q=10^{-4}$ for $q=23.9569$. The simulation results are presented below in Figures 4-6.

In the Figure 4, we illustrate the super-resonance effect for the non-absorbing mesoscale particle with size parameter $q=23.9569$ and refractive index $n_s=1.9$, immersed in air with refractive index of n=1.000241307. These parameters correspond to a resonant mode excited inside the particle with partial wave order $l=34$.

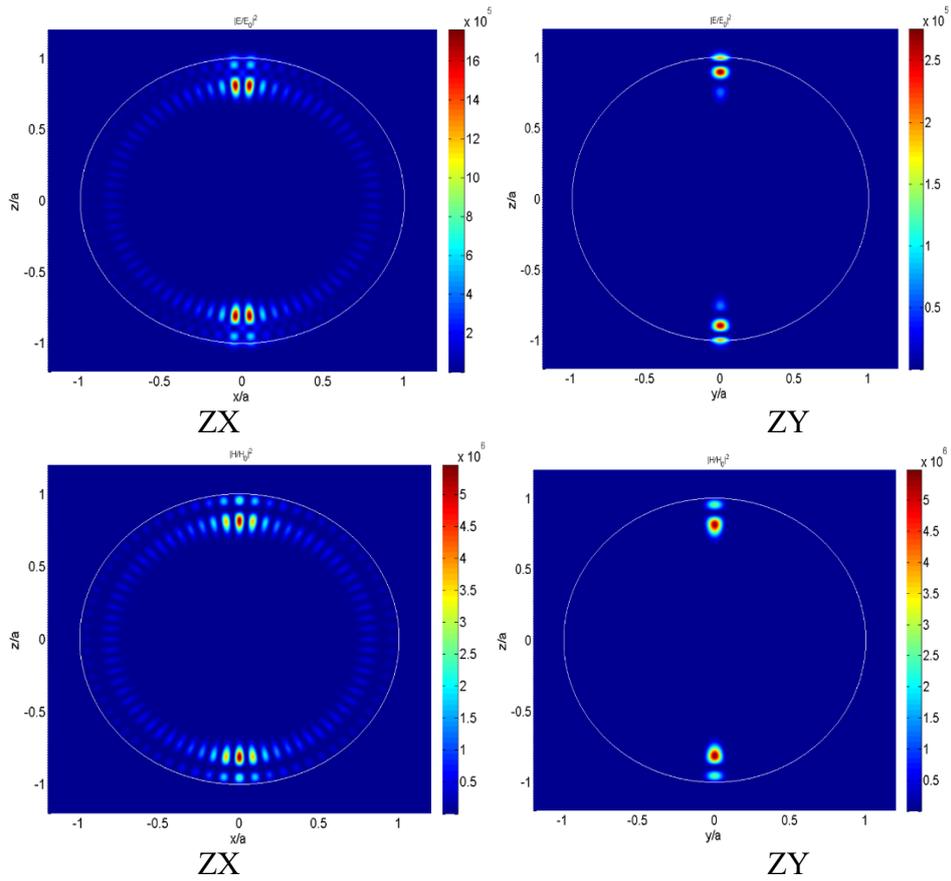

Figure 4. Electric and magnetic hot spots generation for dielectric sphere immersed in air.

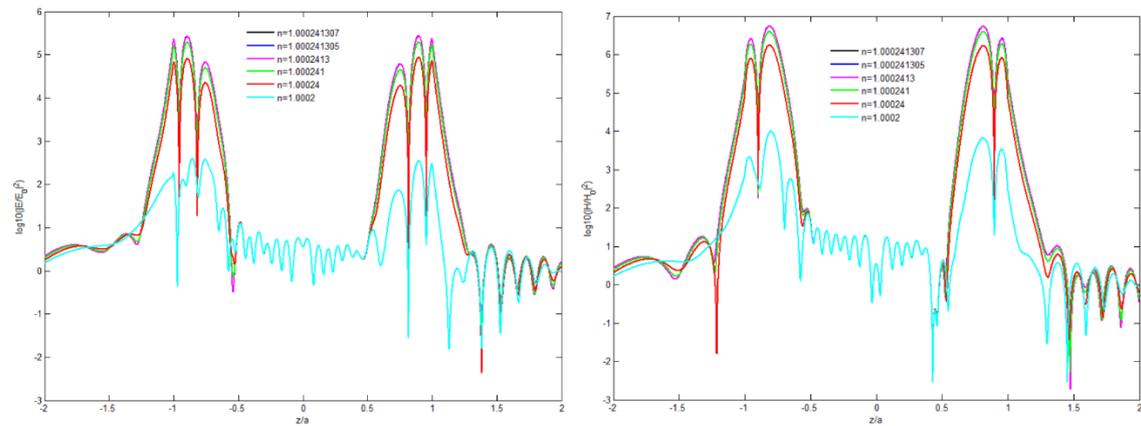

Figure 5. Evolution for electric (left) and magnetic (right) field intensity vs the accuracy of the value of the refractive index of air in the environment.

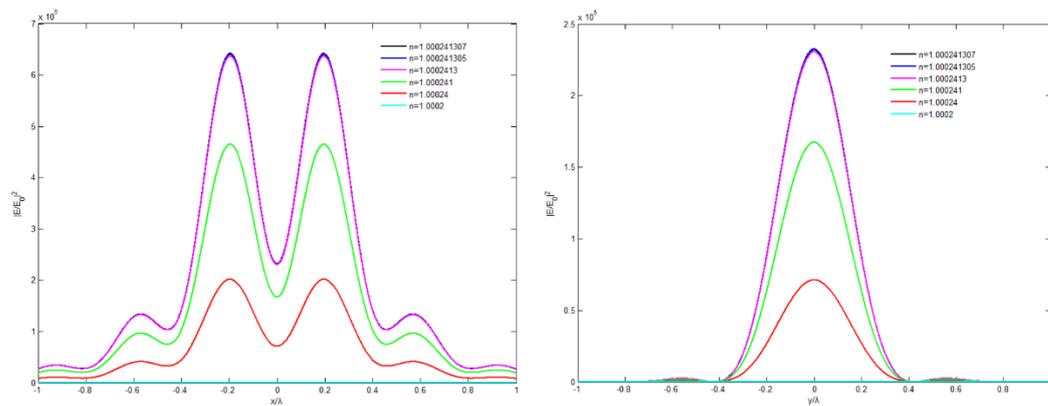

Figure 6. Zoom images of the electric field intensity from Figure 5: Intensity at the bottom of the sphere along x-axis (left) and along y-axis (right).

In the Figures 5-6, we illustrate the super-resonance effect evolution for such dielectric sphere with $q=23.9569$ vs variation of the refractive index of surrounding medium. One can see the changes in refractive index of air up to $\delta n=10^{-7}$ (from $n=1.000241307$ to $n=1.0002413$) hardly noticeable at maximum field intensity at hot spots – from $|E|^2=2.687e5$ to $|E|^2=2.666e5$ and from $|H|^2=5.498e6$ to $|H|^2=5.455e6$, respectively. The changes in refractive index of air from $n=1.000241307$ to $n=1.000241$ leads to change in the field intensity at hot spots almost in 1.4 times – to $|E|^2=1.934e5$ and to $|H|^2=3.988e6$, respectively. Other values of peak intensities are given in Table 2.

Table 2.

| n, air | L, mode | E²(along z-axis) | H²(along z-axis) |
|---|---|---|---|
| 1.000241307 | 34 | 2.687e5 | 5.498e6 |
| 1.000241305 | 34 | 2.681e5 | 5.485e6 |
| 1.0002413 | 34 | 2.666e5 | 5.455e6 |
| 1.000241 | 34 | 1.934e5 | 3.988e6 |
| 1.00024 | 34 | 8.194e4 | 1.736e6 |
| 1.0002 | 34 | 397.8 | 9892 |

This character of the change in the field intensity at hot spots is due to the corresponding behavior of the Mie coefficients. In Figure 7, we show the magnitude of the Mie scattering coefficients $\left|{}^m A_n\right|$ [22] for $1 \leq l \leq 60$ and for different values of $n$. One can see that the resonant scattering coefficient $\left|{}^m A_{34}\right|$ is weakly depends on the change in the refractive index of air (of the medium) when it changes from $n=1.000241307$ to $n=1.0002413$. It could be noted that domain mode index, $l$, decreases with sampling accuracy, from order $l=35$ for $\delta q=10^{-10}$ to order $l=34$ for $\delta q=10^{-4}$.

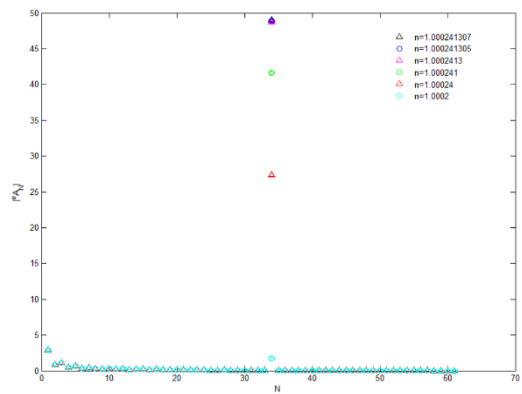

Figure 7. The first 60 Mie $\left|{}^m A_n\right|$ coefficients [22] vs different values of air $n$. Evolution of the partial wave order $l=34$.

The results of comparison the influence of the refractive index of air environment on the super-resonance effect for a sphere with different Mie size parameter sampling accuracy of $\delta q=10^{-4}$ and $\delta q=10^{-10}$ are shown below in Figure 8. In fact, these graphs are an example of calibration curves.

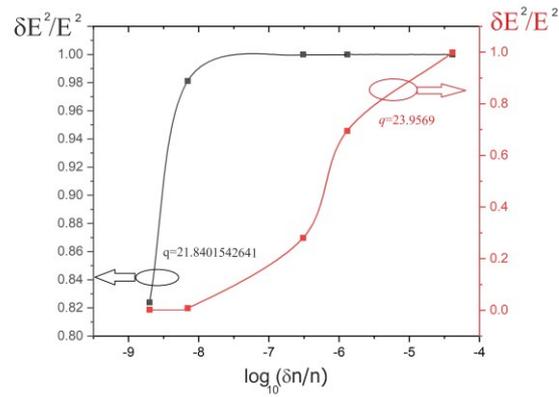

Figure 8. Comparison the influence of the refractive index accuracy of air environment on the super-resonance effect for a sphere with different Mie size parameter sampling accuracy of $\delta q=10^{-4}$ and $\delta q=10^{-10}$.

It can be seen that with the accuracy of the size parameter $\delta q=10^{-10}$ a change in the refractive index of the air medium of about $2*10^{-8}$ is clearly recorded by a change in the intensity of hot spots. At the same time, with a size parameter accuracy of $\delta q=10^{-4}$, a corresponding change in the refractive index at a level of $10^{-6}$ can be registered with a 20% drop in the intensity of hot spots.

**Conclusion**

The refractive index of a material describe a key part of its interaction with electromagnetic waves and is one of the fundamental parameter. Optimization of the main parameters of the particle (Mie size parameter and it accuracy, refractive index of the material of the sphere) and the medium in which it is located will allow the best "tuning" of the sensitivity of the super-resonance effect to changes in the refractive index of the medium. The proposed sensor concept is not limited to air and can be applied to any other medium materials with low losses. Note that this concept does not require working with weak signals (intensities), as well as the use of semi-empirical formulas for calculating the refractive index of the medium.

Considering dependences may also provide a new route for practical selection of mesoscale spheres and surrounding medium to achieve best field location and improved focusing resolution, and a new theoretical background to explain the deep super-resolution observed in microsphere nanoscopy technique [13,28] in super-resonance mode.

**Acknowledgements**

# Supplementary materials

q= 23.9569

$n_1$= 1.000241307

$E^2(zx)$   $E^2(zY)$   $H^2(zx)$   $H^2(zY)$

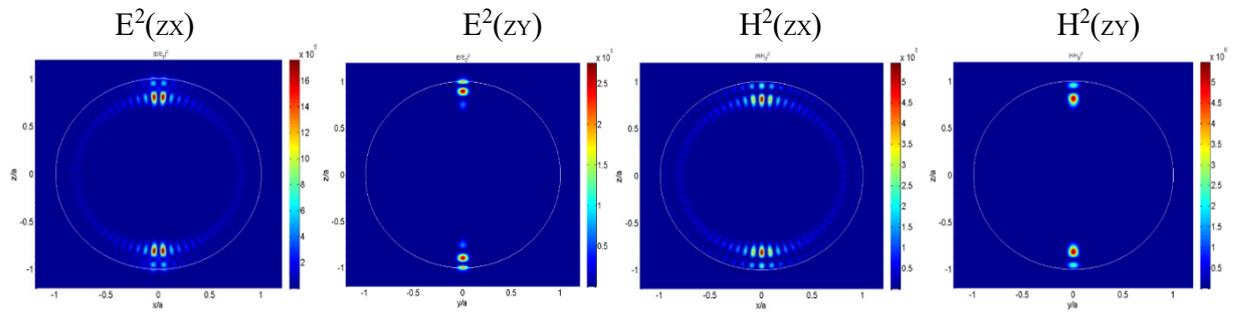

$n_2$= 1.000241305

$E^2(zx)$   $E^2(zY)$   $H^2(zx)$   $H^2(zY)$

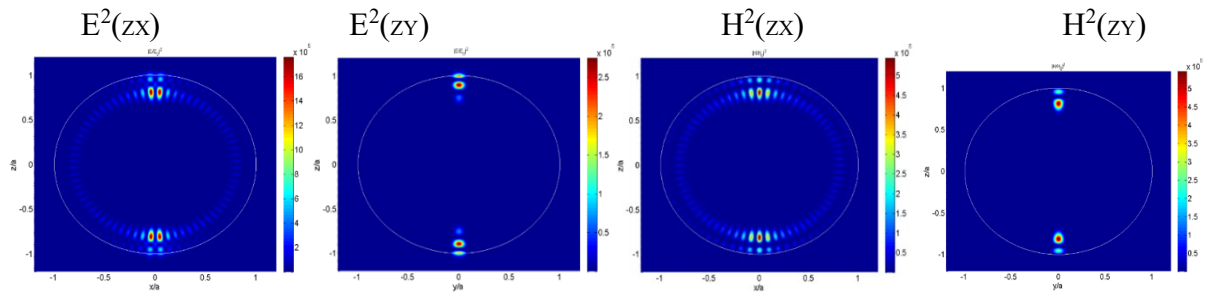

$n_3$= 1.0002413

$E^2(zx)$   $E^2(zY)$   $H^2(zx)$   $H^2(zY)$

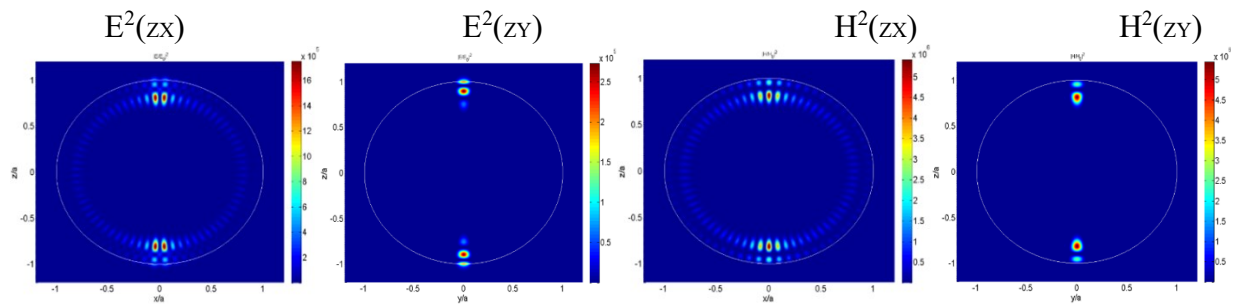

$n_4$= 1.000241

$E^2(zx)$   $E^2(zY)$   $H^2(zx)$   $H^2(zY)$

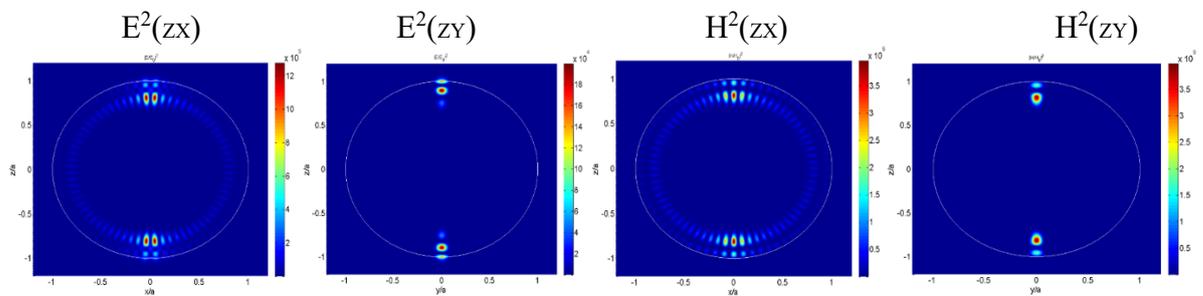

$n_5$= 1.00024

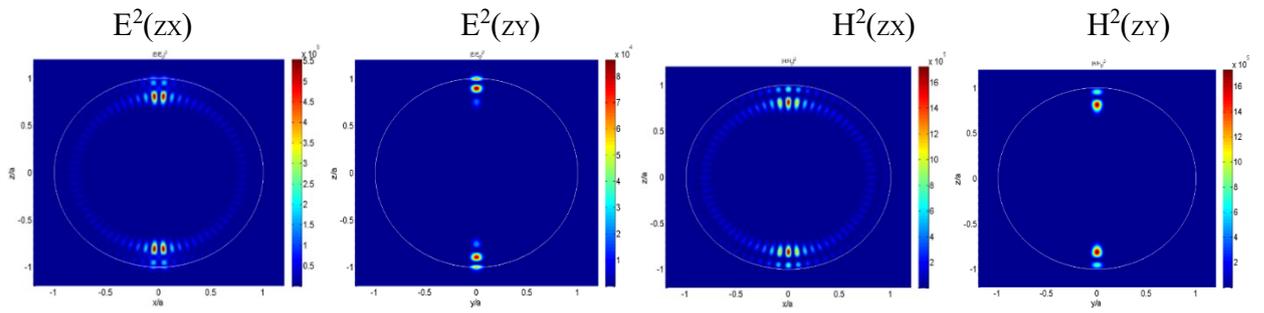

$E^2(zx)$   $E^2(zY)$   $H^2(zx)$   $H^2(zY)$

$n_6$= 1.0002

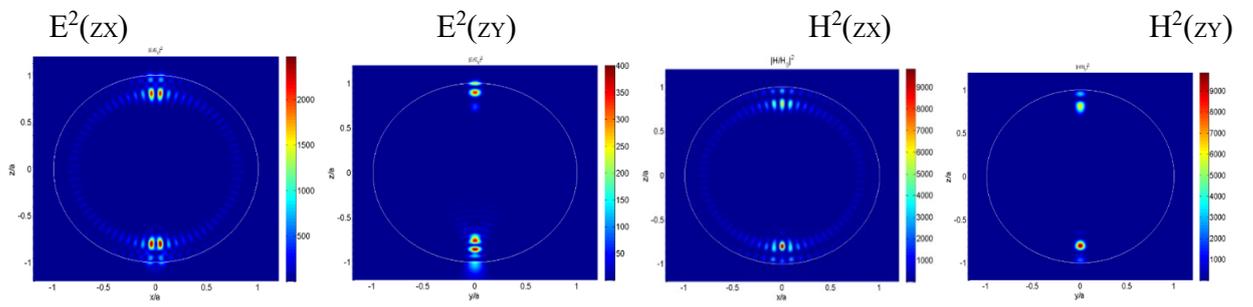

$E^2(zx)$   $E^2(zY)$   $H^2(zx)$   $H^2(zY)$

q=21.8401542641

n₁= 1.000241307

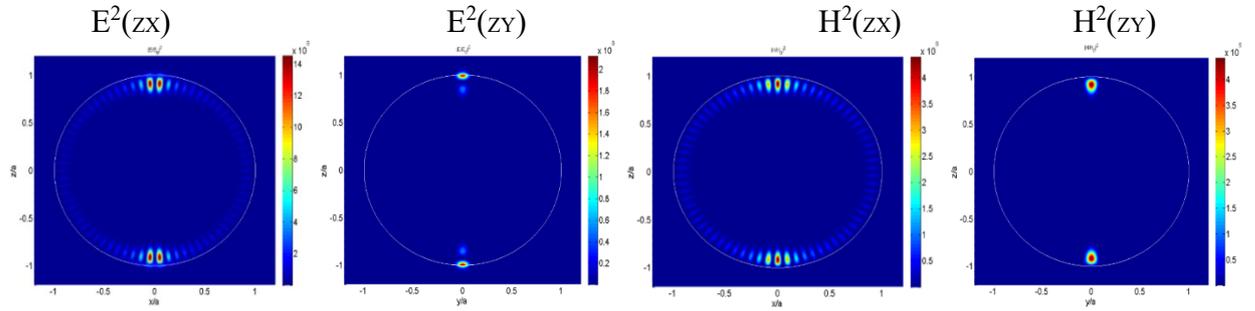

n₂= 1.000241305

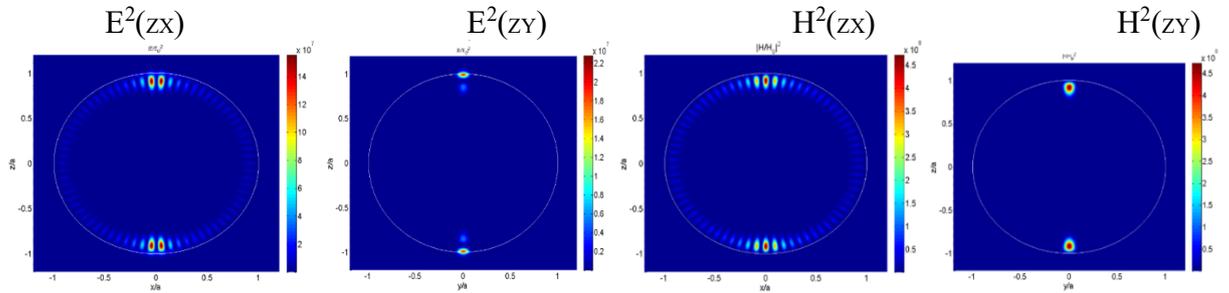

n₃= 1.0002413

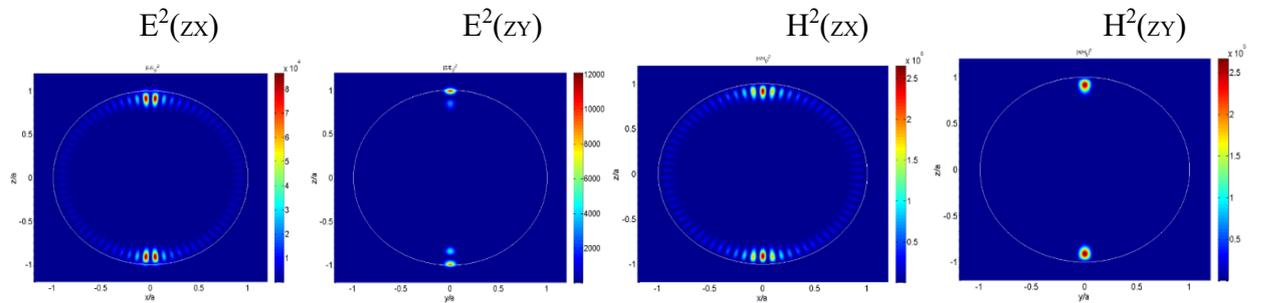

n₄= 1.000241

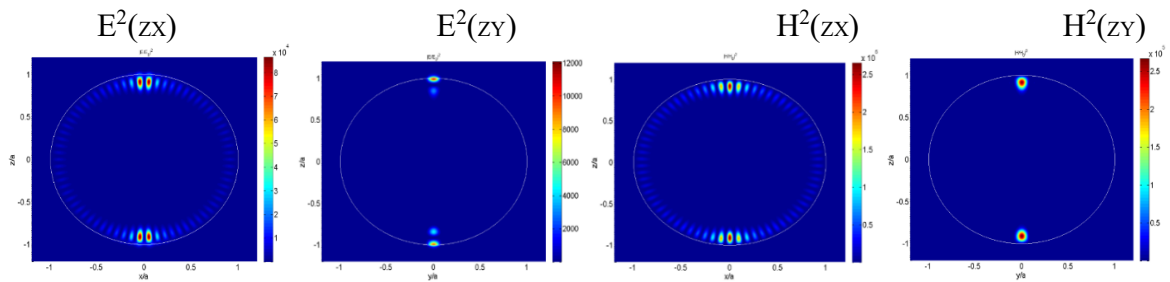

$n_5$= 1.00024

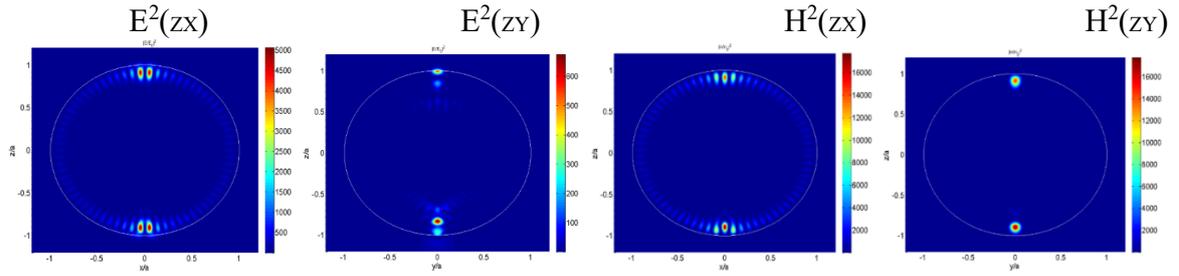

$n_6$= 1.0002

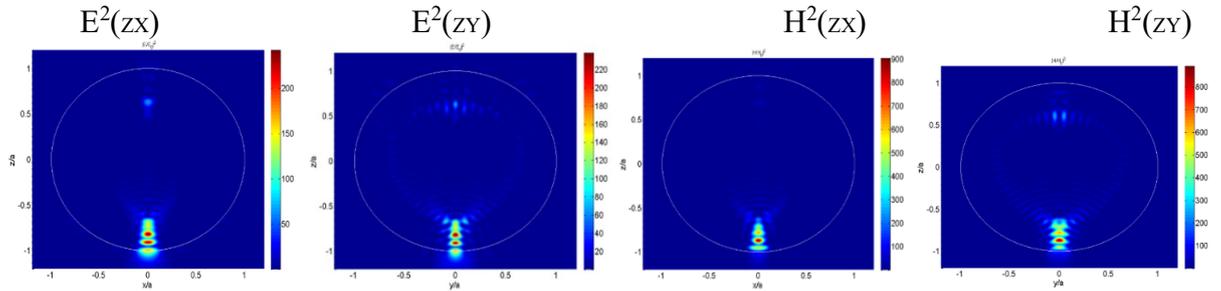

Once super-resonance peaks position and number of modes for selected values of q and refractive index of the medium were found, we simulate the magnetic and electric fields intensity distribution in XZ or YZ plane (incident beam x-polarized, propagates from +z to -z direction). The simulation was performed with spatial resolution a/200 within XZ or YZ plane ranging from -1.2a to 1.2a, where a is radius of particle.